# Engaging with mental health: a global challenge

**David Coyle**
The Computer Laboratory
University of Cambridge
William Gates Building
15 JJ Thomson Avenue
Cambridge CB3 0FD, UK
coyledt@tcd.ie

**Mark Matthews**
Student Counselling Service
Trinity College Dublin
College Green
Dublin 2, Ireland
mark.matthews@cs.tcd.ie

**Gavin Doherty**
School of Computer Science and Statistics
Trinity College Dublin
College Green
Dublin 2, Ireland
gavin.doherty@cs.tcd.ie

**John Sharry**
Child and Adolescent Psychiatry
Mater Misericordiae Hospital
Metropolitan Building
James Joyce Street
Dublin 1, Ireland
jsharry@mater.ie

## Abstract
Using the metrics of the World Health Organisation, the Global Burden of Disease Study has found that mental health difficulties are currently the leading cause of disability in developed countries [1]. Projections also indicate that the global burden of mental health difficulties will continue to rise in the coming decades. The human and economic costs of this trend will be substantial. In this paper we discuss how effectively designed interactive systems, developed through collaborative, interdisciplinary efforts, can play a significant role in helping to address this challenge. Our discussion is grounded in a description of four exploratory systems, each of which has undergone initial clinical evaluations. Directions for future research on mental health technologies are also identified.

## ACM Classification Keywords
H.5.m [Information Interfaces and Presentation]: Miscellaneous – interdisciplinary design, mental health

## Introduction
Mental disorders are health conditions defined by the experiencing of severe and distressing psychological symptoms, to the extent that normal functioning is seriously impaired, and some form of help is usually needed for recovery. The US Surgeon General's first report on mental health concluded that (1) the efficacy of mental health treatments is well documented and (2) a range of effective treatments exist for most mental disorders [2]. Unfortunately international studies also conclude that the majority of people experiencing difficulties do not receive appropriate specialist treatment [2, 3]. Research concludes that in the UK mental health has now overtaken unemployment as the nation's most expensive social problem [4].

**An interdisciplinary challenge**
Addressing the challenges of providing more effective mental healthcare (MHC) services will require the concerted efforts of professionals across a range of disciplines. It is likely – indeed necessary – that technology will play a significant role in future service delivery. Coyle et al [5] identifies two broad challenges which interactive systems can help in addressing [5]:

1. Access/capacity constraints: traditional mental health intervention strategies, particularly talk-based strategies, are time and resource intensive. As a result existing services often do not have sufficient capacity to meet the needs of people requiring professional help.

2. Engagement: research suggests that, even when professional help is available, many clients find it difficult to successfully engage with traditional treatment. The level to which clients engage with their treatment, and draw on their own personal resources, is a major factor in the success of interventions.

With several notable exceptions, early research on the use of technology was generally justified on the basis of increased access - e.g. electronic contact as a natural extension of face-to-face dialogue and the computerisation of self-help materials. Increased engagement and actual improvements in the effectiveness of treatment have received less attention [5]. Collaboration between HCI and MHC professionals can help in maximising the effectiveness of new technologies. While MHC professionals have the necessary domain expertise, HCI researchers are experienced in design methodologies and are likely to have a broader knowledge of the potential uses of new technologies. For example the experience of HCI researchers is important given the high cost of systems failures in sensitive interventions. Other ongoing areas of HCI research, such as designing for personal reflection and behaviour change, can play a valuable role in future research on mental health technologies.

**Examples of exploratory systems**
We have primarily focused on the design of technology to support talk-based, psychological approaches to mental health treatment, e.g. psychotherapy. Reviews of previous research on technology in this area are available in [5, 6]. Over the past 7 years we have developed several exploratory systems.

*Personal Investigator*
Personal Investigator (PI) is a 3D computer game designed to support adolescent mental health interventions. It incorporates Solution Focused Therapy, a goal oriented, strengths based intervention approach. PI is used in clinical sessions involving one therapist and one adolescent. It aims to ease the difficulties many adolescents experience in engaging with face-to-face treatments. Two clinical evaluations of PI have been conducted, the first with 4 adolescents, the second with 22 [7]. Results indicate that PI can have a beneficial impact on interventions, supporting improved client-therapist relationships and improved client engagement.

*PlayWrite*
PlayWrite extended the ideas developed with PI. Rather than providing a fixed game, it provides a flexible game template and enables end users – MHC professionals - to create and adapt the content delivered through this template. Using PlayWrite MHC professionals have created 10 games, implementing different theoretical approaches to MHC and targeting a range of difficulties including depression, anxiety and anger management. For example, gNatenborough's Island supports a six week cognitive behavioural therapy (CBT) intervention for depression. Evaluations of this game began in April 2009 and will run until May 2010.

*Mobile Mood Diary*
Therapists, particularly those practicing CBT, often ask clients to complete paper based mood charts, but adherence can be low and they can provide unreliable information. Mobile Mood Diary, a mobile phone and computer based mood chart system, was developed to make recording moods more convenient and reliable. A controlled school study was run studying compliance variation between a paper-based chart and the Mobile Mood Diary [8]. An initial clinical evaluation with 10 adolescent clients has also been completed. Results showed a high level of adherence amongst participating adolescents over a sustained period.

*Mobile My Story*
My Mobile Story makes use of the multimedia capabilities of mobile phones. It allows adolescents to record inter-session thoughts, ideas and feelings in the form of sounds, pictures, videos and text. This content can then be accessed on a computer to construct therapeutically meaningful stories. Existing paper-based therapeutic exercises can be incorporated into the system as 'therapeutic plans', thereby supporting a range of existing approaches. A pilot clinical evaluation with 5 adolescents has been completed, with further evaluations ongoing. Initial feedback suggests that My Mobile Story can help to increase client engagement between sessions and, according to one therapist, can help to *"personalise the content for therapy sessions"*.

**Design challenges and future directions**

*Theories for design and evaluation*
Mental health technologies remain a relatively unexplored design space. Alongside exploring the potential of a wider range of technologies, there is a need for more detailed investigation of theories, from both MHC and HCI, that can help in generating ideas and support us in reasoning about designs. For example mental health frameworks such as the Skilled Helped Model provide a structured overview of a helping intervention. Behaviour change models such as the Trans-Theoretical Model many also prove useful, as could activity theory which has previously been applied to a range of healthcare settings. Applying models for complex medical evaluations may also prove beneficial.

*Strategies for collaborative design*
While collaboration with domain experts and end users is important is many design areas, it has a particular significance in MHC areas. The ethical requirements and stigma associated with mental health mean that designers often face severe restrictions on access to people experiencing difficulties and to situations in which interactive systems will be used [5]. Faced with such constraints HCI practitioners may have to rely on the expertise and insights of MHC professionals. However research in related healthcare areas has shown that the success of interdisciplinary teams is not something that can be taken for granted. We have discussed initial strategies for collaborative design decision making in [9]. Providing a deeper understanding of collaborative design strategies, and of the techniques which support effective interdisciplinary teams, is an important objective for future research.

*Adaptable systems*
Therapists often work with clients from different socio-cultural backgrounds, experiencing a range of disorders, of varying severities. Furthermore, teams of therapists often have different theoretical backgrounds and adopt a variety of eclectic approaches to working with clients. In such situations fixed systems offer limited benefits. Systems where the core capabilities are re-usable across a range of situations, and which can adapted to the needs of end users have been identified as a important future requirement [5]. PlayWrite represents an example of such a system. In this case MHC professionals play an active role in adapting the system. This approach can also help to support collaboration.

*Support for resource allocation strategies*
Targeting of resources and needs based service provision have become important trend in mental health service delivery. Models such as Stepped Care encapsulate this notion, with distinct steps - intensities of treatment - ranging from inpatient treatment, through primary care teams dealing with issues such as mild depression, to General Practitioners supporting initial contact and diagnosis. Designing interactive systems to support these different levels of care, and transitions between them, represents a major challenge for research in this area.

## Citations